\documentstyle[12pt,epsf,epsfig,here]{article}
\def\beq{\begin{equation}}
\def\eeq{\end{equation}}
\def\ea{\begin{eqnarray}}
\def\beaa{\begin{eqnarray*}}
\def\eea{\end{eqnarray}}
\def\eeaa{\end{eqnarray*}}
\def\bq{\begin{quote}}
\def\eq{\end{quote}}
\def\gappeq{\mathrel{\rlap {\raise.5ex\hbox{$>$}}
{\lower.5ex\hbox{$\sim$}}}}

\def\be{\begin{equation}}
\def\ee{\end{equation}}
\def\bea{\begin{eqnarray}}
\def\eea{\end{eqnarray}}

\def\beq{\begin{equation}}
\def\eeq{\end{equation}}
\def\bea{\begin{eqnarray}}
\def\eea{\end{eqnarray}}
\def\bq{\begin{quote}}
\def\eq{\end{quote}}

\def\gappeq{\mathrel{\rlap {\raise.5ex\hbox{$>$}}
{\lower.5ex\hbox{$\sim$}}}}

\def\lappeq{\mathrel{\rlap{\raise.5ex\hbox{$<$}}
{\lower.5ex\hbox{$\sim$}}}}

\def\SM{Standard Model}

\begin{document}
\pagestyle{empty}

\begin{flushright}
CERN-TH/99-162 \\
hep-ph/9906229
\end{flushright}
\vspace*{2cm}
\begin{center}
{\bf THEORETICAL SUMMARY: }\\
{\bf 1999 Electroweak Session of the Rencontres de Moriond}\\
\vspace*{2cm}
{\bf John ELLIS} \\
\vspace*{0.5cm}
Theoretical Physics Division, CERN \\
CH - 1211 Geneva 23 \\
\vspace*{1cm}

ABSTRACT
\end{center}

The following aspects of the electroweak interactions are
discussed, based on presentations here:
the status of the Standard Model, CP violation,
neutrino masses and oscillations, supersymmetry and
models in extra dimensions, and future projects.
Particular emphasis is laid on the tests of CP and CPT
by KTeV and CPLEAR, on the problems of degenerate neutrinos,
on supersymmetric dark matter, on future long-baseline
neutrino beams, and on muon storage rings that may be
used as neutrino factories.
\vspace*{1cm}
\begin{center}
{\it Invited talk presented at the}\\
{\it 1999 Electroweak Session of the Rencontres de Moriond, Les Arcs}\\
{\it March 1999}
\end{center}
\vspace*{1cm}
\begin{flushleft}
CERN-TH/99-162\\
June 1999
\end{flushleft}

\vfill\eject

\setcounter{page}{1}
\pagestyle{plain}

\section{Status of the Standard Model}
The \SM~ is the rock on which all our knowledge of particle 
physics rests. Novel phenomena such as neutrino masses, that cannot be
accommodated within it, are not thought to contradict it, but rather to take us
beyond the \SM. In the first section of this Summary, I review the status of
the electroweak sector of the \SM, that is the most precisely (and
successfully)
tested.  Then I review some of the accelerating progress in flavour physics and
CP violation, where exciting new developments were reported at this meeting and
more are expected soon. Next I discuss in some detail the growing evidence for
neutrino oscillations, which would take us beyond the \SM~into a new world of
lepton flavour physics. This is followed by a review of new developments in
supersymmetry and further beyond the \SM, in particular the possibility
that
there may be one or more extra dimensions appearing at some large distance
scale. Finally, I review how our theoretical questions may be answered by
future projects such as LEP 2000, Run II of the Fermilab Tevatron,
long-baseline neutrino experiments, the LHC, a linear $e^+e^-$ collider, a
neutrino factory and muon colliders.

As we heard at this meeting~\cite{LEPEWWG}, (almost) everything in the
garden is lovely, as far
as precision tests of the \SM ~ are concerned. Previous discrepancies are
gradually eroding: for example, the discrepancy between the LEP and SLC values of
$\sin^2\theta_W$ is now only about 1~1/2 standard deviations~\cite{SLD},
the $b$-quark
coupling $A_b$ is less than 1 standard deviation from the \SM, and even the
famous $R_b$ deviates  by only about 1~1/4 standard deviations. The
biggest
high-energy problem seems to be in the forward-backward asymmetry for $b$ quarks
$A_{FB}^b$, at a level of about two standard deviations, but at least one
fluctuation at this level could be expected in the high-energy data set. 

One of the most significant potential problems may be in atomic-physics parity
violation. The latest determination of the neutral weak charge
in cesium 
yields $Q_W = -72.06(28)(34)$, to be compared with the \SM ~ prediction  $Q_W =
-73.20(13)$~\cite{BW}. As pointed out by the authors, this discrepancy is
not
compatible with a
vacuum-polarization $S$ contribution in a ``model-independent"
parametrization~\cite{Casal}. 

We heard at this meeting of significant progress in measuring $M_W$ in $p\bar p$
collisions: $80.448 \pm 0.062$~GeV~\cite{Lanc} as well as at
LEP: $80.350 \pm 0.056$~GeV~\cite{Riu}. So far, there are
contradictory indications on the
importance of hadronization effects in $W^+W^- \rightarrow q \bar q q \bar q$
final states~\cite{Hapke}. For example, DELPHI reports possible hints of
colour reconnection
effects in the spectra of low-momentum particles and also an indication of
Bose-Einstein correlations between particles from the $W^\pm$, whereas
ALEPH sees neither of these effects. It is in any case clear that
the effects are not as large as in the model~\cite{EG}, which has
not been tuned to match $Z^0$ decays, and hence cannot be used to
estimate systematic errors in the $M_W$ measurement: see also~\cite{GEHW}.
However, putting together the LEP measurements
reported here there {\it is} a difference of 152 $\pm$ 74 MeV between the
measurements
of $M_W$ in $q\bar q q\bar q$ and $q\bar q \ell \nu$ final
states~\cite{LEPEWWG}, rather larger than the hadronization error usually 
quoted. It remains to
be seen whether this discrepancy will turn out to be significant. The current
world average of direct measurements:
\begin{equation}
M_W  = 80.394 \pm 0.042~{\rm GeV}
\label{newMW}
\end{equation}
has a
reduced error that begins to challenge that on the indirect prediction $M_W = $
80.364 $\pm$ 0.029 GeV within the \SM,  and leaves little room for New
Physics,
such as that suggested in some supersymmetric scenarios.

As seen in Fig.~1, the 
current indirect prediction for the Higgs-Boson mass
is~\cite{LEPEWWG,SLD}: 
\beq
m_H = 76^{+79}_{-67}~{\rm GeV}
\label{one}
\eeq
with a 95\% confidence-level upper limit of 235 GeV, if a conservative
error 
$\pm 0.00090$ on $\alpha_{em}(M_Z)$ is adopted. On the other hand, if a
more
aggressive error $\pm$ 0.00036 is adopted, as suggested on the basis of
hadronic
$\tau$ decays and QCD theory~\cite{Davier}, one finds:
\beq
m_H = 93^{+63}_{-54}~{\rm GeV}
\label{two}
\eeq
So near, and yet so far?  Within the uncertainties (\ref{one},\ref{two}), there
are significant prospects for finding the Higgs boson at LEP during the next
couple of years, or it may be found during Run II at the
Tevatron~\cite{LykkenTeV}, or one may
have to wait for several years of LHC running before it is detected.

\begin{figure}
\hglue3.5cm
\epsfig{figure=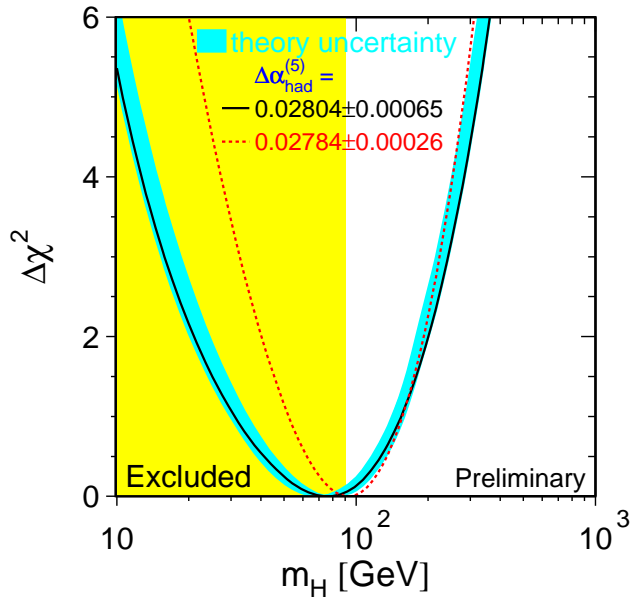,width=8.5cm}
\label{FIG1}
\caption[]{\it The latest result~\cite{LEPEWWG,SLD} 
of fitting precision electroweak data
to obtain an indirect estimate of the mass of the Higgs boson. The most
likely value is very close to the present lower limit from direct searches
at LEP~\cite{Felcini}.}
\end{figure}

In my view, it is a real tragedy that no money can apparently be found to run the
SLC during the year 2000, just when the accelerator has been operating at higher
luminosity than ever, and the SLD detector is poised to make
the most precise measurement of $\sin^2\theta_W$ via $A_{LR}$, as well as
contribute important new information on the $b$-quark
couplings~\cite{SLD}. These
measurements would, in particular, contribute significantly to reducing the
present uncertainty in the prediction of $M_H$, which is one of the main
uncertainties in the LHC enterprise. It seems strange that the equivalent of a
small fraction of the investment in the LHC cannot be found to sharpen this
prediction. 

An important contribution to reducing the uncertainty in $M_H$ is made by those
who measure  $R \equiv \sigma (e^+e^- \rightarrow$ hadrons) $ /  \sigma (e^+e^-
\rightarrow \mu^+\mu^-)$ at intermediate energies~\cite{Qi}, which affects 
the effective  value of $\alpha_{em}$ to be used in analyzing LEP and SLC
measurements via vacuum-polarization diagrams. Uncertainties in $R$ also make
significant contributions to the theoretical error of $\pm$~0.66 ppm in
($g-2)_\mu$, which is large compared to the objective of the new BNL experiment:
$\pm$~0.35 ppm~\cite{g2}. These considerations motivate the important
experimental campaigns
in Beijing and Novosibirsk to remeasure $R$ at center-of-mass energies below 5
GeV~\cite{Qi}. 

Up to what scale can the \SM ~ remain valid?  If $M_H$  is too large, the \SM ~
couplings blow up below the Planck mass $M_P$, and if it is too small, the
effective potential of the \SM ~ is not stable against the development of a
vacuum-expectation value below $M_P$~\cite{invalid}. The range preferred
by precision
electroweak measurements (\ref{one}), (\ref{two}) is still compatible with the
\SM~ remaining valid up to $M_P$, but may be hinting that New Physics  will
appear before then. Some possibilities are discussed in later sections of this
talk.

\section{Flavour and CP}

Until this meeting, all the confirmed CP-violating effects could be explained in
terms of superweak CP violation in the $K^0\bar K^0$ mass
matrix~\cite{superweak}:
\beq
{\rm Im} M_{\Delta s = 2} \not= 0 \rightarrow \epsilon
\label{three}
\eeq
However, we are now sure that there is also CP violation in the
$K^0\rightarrow 2\pi$
decay amplitudes~\cite{KTeV}:
\begin{equation}
\epsilon^\prime / \epsilon = (28.0 \pm 4.1) \times 10^4
\label{KTeVnew}
\end{equation}
In the \SM ~ with six quarks~\cite{KM}, in which there is a CP-violating
phase in the
$W^\pm$ couplings, a non-zero value was to be expected~\cite{EGN}. If
there is only one Higgs doublet, there is no other
CP
violation in the \SM~\cite{EGN}. However, there are additional
CP-violating phases in
extensions of the \SM, such as the Minimal
Supersymmetric Extension of the
Standard Model (MSSM), which may contain arbitrary phases in the soft
supersymmetry breaking mass terms~\cite{Wagner}:
\beq
m_{12}^{*2} A_f \mu, \;\; m_{12}^{*2} M_a \mu
\label{four}
\eeq
Phenomenologically, the most important may be
those associated with the third generation.
There are even more possibilities in general two-Higgs doublet
models~\cite{Kalinowski}.

Confirmation of direct CP violation in the $K^0\bar K^0$ system has come
twenty-three years after it was first calculated~\cite{EGN}.
When we first calculated it, using what  we subsequently
baptized penguin diagrams~\cite{EGNR}, we estimated 
\beq
\vert {\epsilon^\prime\over \epsilon} \vert \lappeq {1\over 450}
\label{five}
\eeq
which is compatible with the value found by NA31~\cite{NA31} and now
confirmed by KTeV (\ref{KTeVnew})~\cite{KTeV}. This
is, of course, pure coincidence, since there have been many conceptual
advances
and calculational improvements since our first calculation:
see, e.g.,~\cite{GW,FR}. The best available recent estimate
is~\cite{latest}:
\beq
{\rm Re} ({\epsilon^\prime\over\epsilon}) = {\rm Im} \lambda_t \left\{ -1.35 +
\left({150~{\rm MeV}\over m_s (m_c)}\right)^2 \times [1.1 R_{SD} B_6^{(1/2)} +
(1.0-0.67 R_{SD}) B_8^{(3/2)}]\right \}
\label{six}
\eeq
where Im$\lambda_t$ is a Kobayashi-Maskawa factor that is expected to lie between 
1.0 and $1.7 \times 10^{-4}$, 
$R_{SD}$ is a short-distance factor that is expected to lie between 7.5 and 8.5,
and $B_6^{(1/2)},B_8^{(3/2)}$ are hadronic matrix elements that are estimated to
lie in the ranges 
\beq
0.8 \lappeq B_6^{(1/2)} \lappeq 1.3~,~~~
0.6 \lappeq B_8^{(3/2)} \lappeq 1.0
\label{seven}
\eeq
It is possible to fit~\cite{latest} the  NA31/KTeV experimental value of 
$\epsilon^\prime/\epsilon$ if one pulls on all these factors,
in particular with a
relatively small strange-quark mass:
\beq
m_s(m_c) \lappeq 126~{\rm MeV} \leftrightarrow 
m_s(2~{\rm GeV}) \lappeq 110~{\rm MeV} 
\label{eight}
\eeq
This is  compatible with other estimates of $m_s$, including 
those from the lattice~\cite{Lellouch} and a recent
determination from $\tau\rightarrow (K n \pi ) \nu$ decays~\cite{Videau}: 
\beq
m_c(m_\tau) = 176^{+46}_{-57}~{\rm MeV}
\label{nine}
\eeq
We may therefore conclude that the NA31/KTeV measurement~\cite{NA31,KTeV} 
of
$\epsilon^\prime/\epsilon$ {\it does not require physics beyond the Standard
Model}. 
On the other hand, there is room for a supersymmetric
contribution~\cite{sepsilon}, e.g., via
an enhanced $Z \bar d s$
vertex~\cite{Silvestrini}, which could
enhance significantly the rates for some rare $K^{0,\pm }$ decays:
\bea
B(K_L^0 \rightarrow \pi^0 \nu\bar\nu) &\sim& 3\times 10^{-11} \rightarrow 
27 \times 10^{-10} \nonumber \\
B(K_L^0 \rightarrow \pi^0 e^+e^-)_{direct} &\sim& 5\times 10^{-12} \rightarrow
43\times 10^{-11} \nonumber \\
B(K^+ \rightarrow \pi^+\nu\bar\nu) &\sim& 8\times 10^{-11} \rightarrow 8\times
10^{-10}
\label{ten}
\eea
putting them within reach of forthcoming experiments. 

We can expect in the near future a new measurement of  $\epsilon^\prime /
\epsilon$ from the NA48 experiment~\cite{Mikulec}, as well as further data
from KTeV and
subsequently from the KLOE experiment at DA$\phi$NE, 
perhaps closing an exciting chapter
in $K$ physics. What are the next important steps in $K$ decays? The most
interesting
decay modes may be those mentioned above:
$K_L\rightarrow \pi^0\bar\nu\nu$ violates CP and could occur at a rate much
larger than that predicted by the \SM~\cite{EGN}, 
$K^+\rightarrow \pi^+\bar\nu\nu$ could provide a good measurement of the
Cabibbo-Kobayashi-Maskawa mixing angles, and
$K_L^0\rightarrow \pi^0e^+e^-$ may be dominated by direct CP violation. It is
surely worth pursuing these decays in parallel with the large investment
currently being made in B physics, via  HERA-B, CLEO, BaBar, BELLE, CDF,
D0,
BTeV and LHCb. There are interesting prospects for  rare $K$ decay
experiments at
FNAL, BNL, KEK and CERN. Can the world community agree on at least one
next-generation rare $K$ decay experiment?

There was active discussion here of T and CPT violation. The discovery of CP
violation in the $K^0 \bar K^0$ mass matrix in 1964 meant that either CPT
or T
should be violated, or both. Since CPT violation is not permitted by quantum
field theory, 	it is generally expected to be absent. However, the possibility of
CPT violation has been raised in the context of quantum
gravity~\cite{CPTQG}, and it is good
that experiments continue to search for it.  It is known that $K_L^0\rightarrow
2\pi$ decay  is not due to CPT violation, and therefore one expects that the
observed CP violation should be accompanied by T violation at the same rate,  but
this was not observed directly until recently. The CPLEAR collaboration reported
here~\cite{CPLEAR} a measurement of the asymmetry 
\beq
{P_{K\rightarrow \bar K} B(K\rightarrow \pi^-e^+\nu) -
P_{\bar K\rightarrow  K} B(\bar K\rightarrow \pi^+e^-\bar\nu) \over
P_{K\rightarrow \bar K} B(K\rightarrow \pi^-e^+\nu) +
P_{\bar K\rightarrow  K} B(\bar K\rightarrow \pi^+e^-\bar\nu)}
\label{eleven}
\eeq
This is a direct probe of reciprocity and hence T violation if 
$B(K\rightarrow \pi^-e^+\nu) = B(\bar K\rightarrow \pi^+e^-\bar \nu)$, which has
indeed been checked by independent CPLEAR measurements~\cite{Filipcic}.
Thus the CPLEAR
collaboration has indeed observed T violation~\cite{EM,Lola}.  The KTeV
collaboration reported
here a beautiful measurement of a T-odd asymmetry in $K^0_L\rightarrow
\pi^+\pi^-e^+e^-$~\cite{Yamanaki}: since this is not a direct test of
reciprocity, this could be
due to either T violation or CPT violation or both~\cite{EM},  at least
until more
experimental information is available.

Another exciting development at this meeting was the report of a
two-standard-deviation CP-violating asymmetry in $B^0\rightarrow J/\psi
K_S^0$~\cite{Schmidt}. In
the absence of other information, this could in principle be due
to a superweak
effect, so we need another decay mode such as $B^0\rightarrow \pi^+\pi^-$ to
confirm its interpretation.  However, we cannot avoid noticing that its sign and
large magnitude agree with the predictions of the Kobayashi-Maskawa
model~\cite{KM}, as seen in Fig.~2: in
particular, the sign of $\epsilon_K$ agrees with the apparent sign of
$\sin 2\beta$. It is encouraging  that the values of the
Cabibbo-Kobayashi-Maskawa parameters extracted from the data are consistent with
na\"\i ve model predictions~\cite{Romanino}: 
\bea
{\lambda\over c}~\sqrt{(1-\bar\rho)^2 + \bar\eta^2} &=& \left\vert{V_{td}\over
V_{ts}}\right\vert = \sqrt{m_d\over m_s} \nonumber \\
{\lambda\over c}~\sqrt{\bar\rho^2 + \bar\eta^2}  &=&
\left\vert{V_{ub}\over V_{cb}}\right\vert = \sqrt{m_u\over m_c} 
\label{twelve}
\eea
where $c \equiv \sqrt{1 - \lambda^2}$. These model predictions and global
Cabibbo-Kobayashi-Maskawa
fits~\cite{Ali} agree in
predicting that the $B_s^0 - \bar B_s^0$  mixing parameter $\Delta m_s$ should be
observable ``soon". Indeed, both SLD~\cite{Moore} and the LEP
collaborations~\cite{Bloch} reported here not
only interesting lower limits on $\Delta m_s$, but also enticing hints that it
may lie just around the corner. 

\begin{figure}
\hglue3.5cm
\epsfig{figure=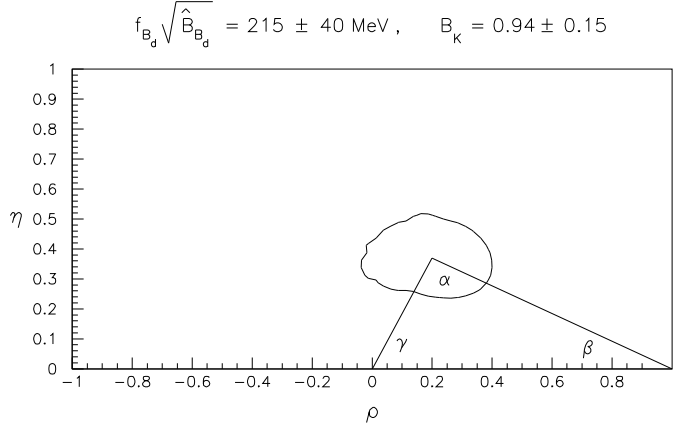,width=10cm}
\label{FIG2}
\caption[]{\it A recent global fit to data on charged-current interactions
and $\epsilon_K$~\cite{Ali}, which suggests a value of sin$2\beta$ within
the range
reported by CDF~\cite{Schmidt}.}
\end{figure}

\section{Neutrino Masses}

These are known to be much smaller than those of the corresponding charged
leptons:
\bea
m_{\nu_e}~ \lappeq ~~2.5~{\rm eV} &\ll& m_e \sim 1/2 ~~{\rm MeV}\nonumber \\
m_{\nu_\mu} \lappeq 160~{\rm keV} &\ll& m_\mu \sim 100 ~~{\rm MeV}\nonumber \\
m_{\nu_\tau} \lappeq~ 15~{\rm keV} &\ll& m_\tau \sim 1.78 ~{\rm MeV}
\label{thirteen}
\eea
so one might be tempted to suspect that they vanish. However, we have learnt that
particles have zero mass (e.g., the photon and gluon) only if they are associated
with an exact gauge symmetry (e.g., the U(1) of QED and the SU(3) of QCD) and a
corresponding conserved charge (e.g., Q$_{\rm em}$ and colour). There is
no
candidate gauge symmetry available to conserve lepton number $L$, and GUTs
generically predict non-zero neutrino masses. 

It is possible for these to appear
even if there are no new particles beyond the Standard Model, e.g., via a
non-renormalizable interaction~\cite{BEG} of the form
\beq
{(\nu_LH)~(\nu_LH)\over M}
\label{forteen}
\eeq
but the most plausible hypothesis is that $M$ represents the mass scale of some
new renormalizable gauge theory, such as the GUT mass scale. Interactions of the
form (\ref{forteen}) arise from the exchange of massive singlet fermions (often
called ``right-handed neutrinos" $\nu_R$), and most models of neutrino
masses are based on
the see-saw form of mass matrix that mixes then with conventional left-handed
neutrinos~\cite{seesaw}:
\beq
(\nu_L~,~~\nu_R)~~\left(\matrix{0&m \cr m&M}\right)~~\left(\matrix{\nu_L\cr
\nu_R}\right)
\label{fifteen}
\eeq
where the off-diagonal ``Dirac" elements $m$ are generally comparable to
conventional quark and lepton masses whereas $M \gg m_W$, leading to light
neutrino masses
$m_\nu
\sim m^2/M$. As an example of possible orders of magnitude, with $m \simeq$ 10
GeV and $M \simeq 10^{13}$ GeV one finds $m_\nu \sim 10^{-2}$ eV.

Within this general framework, many models are possible. Could there be other
light neutrinos? Neutrino counting at LEP~\cite{LEPEWWG} tells us that
these could only be sterile $\nu_s$. But then
what forbids a large gauge-invariant mass term $M_s (\nu_s\nu_s)$
with $M_s \gg M_W$?  In my view,
this is a major objection to models with light sterile neutrinos and/or
right-handed neutrinos. On the basis of (\ref{fifteen}), most theorists expect
the light neutrinos to be predominantly left-handed and to have effective
Majorana masses $m_{eff}(\nu_L\nu_L)$. It often used to be thought that neutrino
mixing would be small, by analogy with the small quark mixing angles. However,
now it is widely recognized that this need not be the
case~\cite{Allanach}. For one thing, the
Dirac masses might not be directly related to conventional quark or lepton
masses, and, for another, the heavy Majorana mass $M$ has no good reason to be
diagonal in a basis aligned closely with the light charged-lepton
flavours~\cite{ELLN}.

	It is widely thought that oscillations between neutrinos with different masses
may explain the anomalies seen in solar and atmospheric neutrinos. In the solar
case, the Standard Solar Model is nicely consistent with the powerful constraints
of helioseismology, and models with large mixing of solar material also seem to be
disfavoured~\cite{Haxton}, so it seems that neutrino physics must be the
culprit for the
persistent solar-neutrino deficit. In the case of atmospheric neutrinos, it has
been shown that neutrino decays and modifications of special and general
relativity do not fit the data well~\cite{nofit}.

Among the indications of neutrino oscillations, the evidence for
$\bar\nu_\mu\rightarrow\bar\nu_e$ oscillations presented by
LSND~\cite{Mills} has not been
confirmed by KARMEN~\cite{Jannakos}, though it is not excluded,
either. The MiniBooNE
experiment at FNAL should confirm or refute this signal, as well as the MINOS
long-baseline experiment discussed later.

As we heard at this meeting~\cite{Casper}, the range of $
\Delta m^2$ favoured by the Super-Kamiokande atmospheric-neutrino
data~\cite{SuperK} has
recently been elevated, as seen in Fig.~3. This is a result
of higher statistics for
contained
events, an improved Monte Carlo including a more complete treatment of the
Earth's magnetic field (that is validated by its agreement with the observed
East-West effect~\cite{EW}), and combining the contained events with data
on up-going
muons. The Super-Kamiokande evidence is supported by data from
MACRO~\cite{MACRO} and Soudan II~\cite{Soudan}. It remains important to
reduce the systematic uncertainties in the
interpretation of the data~\cite{Lipari}. This will require, e.g., more
precise measurements of
the primary cosmic-ray spectrum. New data tend to favour lower fluxes than used in
some Monte Carlos: the AMS space experiment may be able to
help here. Better
measurements of the differential cross-sections
$d^2\sigma^{\pi,K}/dx_{\parallel} dp_\bot$ for secondary-particle production are
also needed: here is a possible role for a new experiment at the CERN PS
accelerator. Also needed is a fully three-dimensional cosmic-ray Monte Carlo:
there are hints that this might further elevate the preferred range of $\Delta
m^2$. Measurements of cosmic-ray secondaries on balloon
flights~\cite{balloon} could be helpful
in constraining and validating such Monte Carlos. It should also not be forgotten
that aspects of the interpretation depend on low-energy $\nu$ cross-section
measurements~\cite{Marteau}, both total and exclusive: here the close-up
detectors in the K2K
and NuMI beam lines will play useful roles, as might measurements by CHORUS
and/or NOMAD.

\begin{figure}
\hglue3.5cm
\epsfig{figure=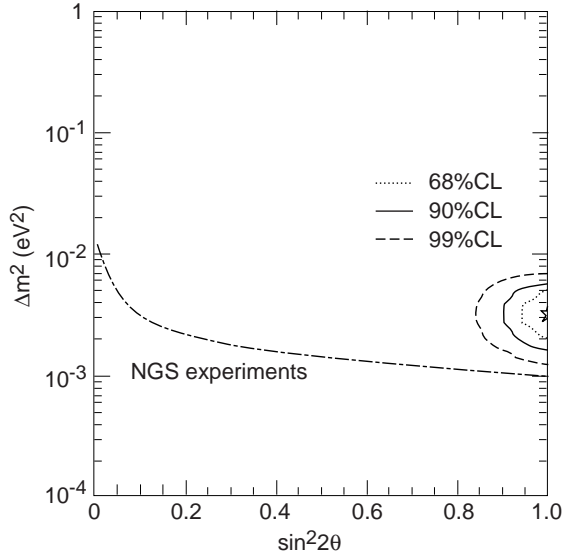,width=7.5cm}
\label{FIG3}
\caption[]{\it The range of parameters for $\nu_{\mu} - \nu_{\tau}$
oscillations preferred by the Super-Kamiokande data at the indicated
levels of confidence~\cite{Casper}, compared with the estimated sensitivity 
to $\tau$ appearance of either the OPERA or the ICARUS experiment in the
CERN-Gran Sasso long-baseline beam.}
\end{figure}

In the case of solar neutrinos, there are  three regions of neutrino-mixing
parameters that are compatible with the rates observed in the different
experiments~\cite{Smirnov}: Kamiokande and Super-Kamiokande at high
energies, Homestake that is
also sensitive to Be neutrinos, and SAGE and GALLEX that are also sensitive to
$pp$ neutrinos. The favoured parameter regions are the small-angle 
Mikheyev-Smirnov-Wolfenstein~\cite{MSW} (MSW) (SMA)
solution with
$\Delta m^2 \sim 10^{-5}$ eV$^2$ and
$\sin^22\theta\sim 10^{-2}$ to $10^{-3}$, the large-angle MSW (LMA) solution with
$\Delta m^2 \sim 10^{-4}$ to $10^{-5}$ eV$^2$ and $\sin^22\theta\sim 1$, and the
vacuum-oscillation (VO) solution with $\Delta m^2 \sim 10^{-10}$ eV$^2$ and~$\sin^22\theta\sim 1$.

In the case of atmospheric neutrinos~\cite{Smirnov}, the possibility of
large
$\nu_\mu\rightarrow\nu_e$ oscillations is excluded by
Chooz~\cite{Nicolo}~\footnote{More data are on their way from the
Palo Verde
experiment~\cite{Wang}.} as well as by
Super-Kamiokande itself. However, $\nu_\mu \rightarrow \nu_e$ oscillations could
still be present at a subdominant level, and exploring this possibility is a key
task for future experiments. The dominant $\nu_\mu$ oscillations may be into
either $\nu_\tau$ or a sterile neutrino $\nu_s$. One way to distinguish these
possibilities may be via $\pi^0$ production by atmospheric
neutrinos~\cite{Smirnov}: they can be
produced by $\nu_\tau$ interactions, but not by $\nu_s$. Another signature is in
the zenith-angle distribution of upward-going muons~\cite{Casper}. In both
cases,
Super-Kamiokande data may be starting to favour $\nu_\mu\rightarrow\nu_\tau$
oscillations. A key test will be the neutral- to charged-current ratio in
long-baseline experiments. However, for me the crucial experiment is to observe
$\tau$ production by either atmospheric or long-baseline neutrinos,
as discussed later.

In the case of solar neutrinos, three measurements may soon be able to
discriminate between the proposed SMA, LMA and VO
solutions~\cite{Smirnov}. Super-Kamiokande
reports a distortion of the recoil $e$ energy spectrum~\cite{Casper}, in
good agreement with
that expected in the VO solution. The SMA solution generally predicts a smaller
distortion (and the LMA solution an even smaller distortion). However, this
difficulty may be overcome if the high-energy hep neutrino production
cross-section, which is difficult to calculate reliably, is (much) larger than
the value assumed in Standard Solar Model calculations~\cite{bighep}. Both
the SMA and LMA
solutions tend to predict some day-night difference, but not the VO solution. The
present Super-Kamiokande data constrain the possible magnitude of this day-night
effect, thereby constraining the SMA and LMA parameter regions. If any day-night
effect were to be seen, its dependence on the solar nadir angle could in
principle discriminate further between different parameter
choices~\cite{nadir}. Some seasonal
variation is predicted in all models, because of the trivial geometric effect due
to the eccentricity of the Earth's orbit. However, a larger seasonal variation is
predicted in VO solutions, and this might be further enhanced at high energies.

Oscillation experiments can determine differences in neutrino masses squared, but
not the absolute values of their masses. Astrophysical and cosmological data
already exclude $m_\nu \gappeq $ 3~eV~\cite{cosmo}, and future data may be
sensitive to
$m_\nu\gappeq$ 0.3 eV~\cite{HET}~\footnote{Negative results of
searches, using the CERN beam, for oscillations involving neutrinos with
masses of interest to cosmology were also
reported here~\cite{Messina,Salvatore}.}. Below this range, neutrinos
would not
constitute a
significant fraction of the astrophysical dark matter, though masses in the range
$\gappeq$ 0.03 eV favoured by atmospheric-neutrino data would have some
cosmological implications~\cite{HET}. 

Are there any experimental or theoretical reasons why the three flavours of
neutrinos should not all be almost degenerate with $\bar m \gappeq$ 2 eV ? There
is a strong constraint from the absence of neutrinoless double-$\beta$
decay~\cite{Klap}:
\beq
<m_\nu> \simeq \bar m \times \left\vert c^2_2~c^2_3~e^{i\phi} +
s^2_2~c^2_3~e^{i\phi^\prime} + s^2_3~e^{2i\phi^{\prime\prime}} \right\vert \lappeq
0.2~{\rm eV}
\label{sixteen}
\eeq
Here $\nu_\mu\rightarrow\nu_e$ oscillations suggest that the last term in
(\ref{sixteen})  may be neglected, in which case the first two terms would need
to cancel at the 90 \% level. Thus $\phi^\prime \simeq \phi +\pi$ and
\beq
c^2_2 - s^2_2 = \cos 2\theta_2 \lappeq 0.1
\label{seventeen}
\eeq
implying that $\sin^2 2\theta_2 \gappeq 0.99$. This certainly excludes the SMA
solution, and very likely LMA solutions, but not VO solutions. However, in the
latter case, the neutrino mass degeneracy would need to be accurate to one part in
10$^{10}$ ! The neutrino-mass degeneracy would be broken by non-universal
loop
corrections: we find~\cite{EL} that these are much larger than would be
allowed in the VO
solution, that in the MSSM they have the wrong sign for the LMA solution, and
that in one favoured neutrino-mass texture they produce unacceptable patterns of
mixing angles. We conclude that degenerate neutrinos are distinctly
problematic~\cite{others}.

Some cute neutrino calculations were reported here. One was of the 
related processes
$\bar\nu\nu\rightarrow\gamma\gamma\gamma$,
$\gamma\gamma\rightarrow\gamma\gamma\gamma$ and
$\gamma\gamma\rightarrow\bar\nu\nu\gamma$~\cite{Matias}. These turn out to
have cross-sections
9 to 13 orders of magnitude larger than the corresponding 2$\rightarrow$ 2
processes $\bar\nu\nu\rightarrow\gamma\gamma$, $\gamma\nu\rightarrow\gamma\nu$
and $\gamma\gamma\rightarrow\bar\nu\nu$! A full Standard Model calculation
agrees
with an effective Lagrangian result at low energies $E \ll m_e$. One of the
consequences is that the neutrino mean free path in a supernova becomes smaller
than the core radius, with potentially important consequences for supernova
simulation codes, that are now being revised to incorporate these effects. The
effects of multibody neutrino exchange in neutron stars was also
discussed~\cite{Tytgat}. A
first estimate had overestimated its importance by 63 orders of magnitude! The
new calculations indicate that a neutrino condensate forms in a neutron star,
 changing its mass by about 30 kg out of 10$^{30}$ kg. By comparison, two-body
neutrino effects change the mass by about 10$^{14}$ kg.

\section{Supersymmetry and Further Beyond}

The phenomenological motivation for supersymmetry is very simple, namely to
stabilize the gauge hierarchy. The theory of supersymmetry is not too complicated
either, since all the sparticle couplings are related to the gauge and Yukawa
couplings of the \SM. However, more complications arise when one considers the
soft supersymmetry breaking needed to elevate sparticle masses. These are usually
parametrized as
\beq
(m^2_0)^j_i~\phi^i~\phi_j~,~~~M_a \tilde V_a \tilde V_a~,~~A^{ijk}\lambda^{ijk}
\phi_i~\phi_j~\phi_k~,~~ B^{ij}~\mu^{ij}~\phi_i~\phi_j
\label{eighteen}
\eeq
which are usually quoted as leading to 105 free parameters
in the MSSM. These are often
simplified by imposing universality at some input GUT or string scale:
\beq
M_a = m_{1/2}~,~~ (m^2_0)^j_i = m^2_0~\delta^j_i~,~~ A^{ijk} = A~,~~ B^{ij} = B
\label{nineteen}
\eeq
before renormalization. Universality for the gaugino masses $M_a$ may be more
plausible than that for the scalar masses, which is particularly questionable for
Higgs scalar masses. 

However, one should also bear in mind the possible appearance of
other supersymmetry-breaking parameters~\cite{Jack} corresponding to
interactions of the form
\beq
\phi^{i}~\phi_j~\phi_k~,~~\varphi_i~\varphi_j
\label{twenty}
\eeq
which should be considered equally soft if they do not generate quadratic quantum
divergences. Important questions to understand include whether string/$M$ theory
can generate terms like (\ref{twenty}), and what might be their implications for
phenomenology, particularly in the $\tilde t, \tilde b$ and Higgs sectors. When
considering the phenomenology of the MSSM, it is important to include the
constraint of correct electroweak symmetry breaking (as characterized by $m_Z$
and $\tan\beta$) after renormalization, and also to check that there is no
lower-energy charge- and/or colour-breaking vacuum. These issues, together with
the implementation of experimental constraints on $m_h, m_A$ and sparticle
masses, are discussed in more detail later.

	The phenomenological signatures of supersymmetry to be sought by experiments
vary in different theoretical scenarios. One basic issue is whether $R$ parity is
conserved or not. In the latter case, one should look for sparticle decays into
quarks and/or leptons, providing the opportunity of looking for bumps in the
invariant masses of leptons and/or jets~\cite{Arnoud}. On the other hand,
if $R$ parity is
conserved, the lightest supersymmetric particle (LSP) is stable, and hence
expected to be present in the Universe as a relic from the Big Bang. Upper limits
on the relative abundances of anomalous heavy isotopes of many elements conflict
with calculations of the relic abundance, suggesting that any such relic cannot
bind to conventional matter, and hence should not have electric charge or strong
interactions~\cite{EHNOS}. Thus the LSP is expected to be a neutral
weakly-interacting
particle such as the lightest neutralino
$\chi$. This leads to the ``classic" missing-energy signature of supersymmetry.
On the other hand, if the lightest neutralino is not the LSP, one expects $\chi$
decays to provide characteristic signatures, e.g., $\chi\rightarrow\gamma\tilde
G$ if the gravitino $\tilde G$ is the LSP. This is the generic prediction of
models in which supersymmetry breaking is communicated to the observable sector by
a messenger gauge-interaction sector~\cite{GiuRat}.

Such gauge-mediated models have recently attracted interest for two reasons. One
is theoretical: they provide a natural explanation why supersymmetry-breaking
scalar masses should be generation-independent~\cite{GiuRat}. The second
is experimental: they
might explain the $\bar pp \rightarrow e^+e^-\gamma\gamma E\llap{$/$}_T + X$
event seen by CDF~\cite{CDF}. However, popular supersymmetric
interpretations of this event
have now been almost excluded by searches at LEP as well as the Tevatron 
itself~\cite{Antonelli,Numerotski}.
Indeed, since the lightest neutralino is directly detectable in gauge-mediated
and $R$-breaking models, unlike the standard $R$-conserving model with a
neutralino LSP, the experimental constraints on MSSM parameter space are
generally stronger in these alternative scenarios. In the following, we
concentrate on the ``conservative" $R$-conserving neutralino LSP scenario.

Important constraints on the MSSM parameter space are provided
not only by direct
searches for unstable sparticles such as sleptons and charginos, but also by
unsuccessful searches for the Higgs bosons of the MSSM. These are sensitive to
other sparticle masses via loop corrections, so lower limits on Higgs masses may
be interpreted as lower limits on the universal soft supersymmetry-breaking
masses, particularly if the input Higgs scalar masses are also universal.
Requiring the relic $\chi$ density to be in the range allowed by astrophysics and
cosmology: $0.1 \leq \Omega_\chi h^2 \leq 0.3$ (where $\Omega_\chi \equiv
\rho_\chi / \rho_c$, with $\rho_c$ the critical density and $h$ the Hubble
constant in units of 100 km s$^{-1}$ Mpc$^{-1}$) further constrains the
MSSM
parameter space. In a combined analysis a couple of years ago~\cite{EFOS},
it was found that
$m_\chi \geq$ 40 GeV, to be compared with a lower limit of 20 GeV from LEP
searches for charginos and neutralinos alone.

One might begin to wonder whether the constraints on the parameter space of the
MSSM are still compatible with the phenomenological motivation of avoiding the
need to fine-tune parameters in order to maintain the gauge
hierarchy. A first
attempt to capture the idea of fine tuning was the ``price"~\cite{price}
\beq
\Delta\equiv \max_{i} \Delta_i~:~~\Delta_i \equiv {\partial \ln m^2_Z\over
\partial \ln ~a_i}
\label{twentyone}
\eeq
where the $a_i$ are input MSSM parameters. The LEP data impose the price $\Delta
\geq$ 13~\cite{CEOP}. Is this a lot or a little? There is no objective
reply, and the price
may be reduced significantly by postulating some theoretical correlations between
the MSSM parameters. Alternatively, it was proposed here~\cite{Strumia} 
that one evaluate the
percentage of the {\it a priori} MSSM parameter space that is still permitted by
the experimental constraints. By one estimate, the residual conditional
probability $P$ (LEP$\vert$susy)$ \simeq$ 5 \%. However, what one really wants to
know~\cite{Strumia} is the opposite conditional probability
\beq
P({\rm susy}\vert{\rm LEP}) =  {P({\rm susy})\times 5 \%\over 1-95 \% \times
P({\rm susy})}
\label{twentytwo}
\eeq
which depends on the {\it a priori} probability assignment $P({\rm susy})$.
Clearly the estimate 
$P$(LEP$\vert$susy) $\simeq$~5~\% depends on the choice
of measure in the MSSM parameter space made in specifying (susy). If you are
totally convinced by the original choice made: $P$(susy) = 1, then
(\ref{twentytwo}) tells you that you still believe it: $P$(susy$\vert$LEP)
= 1! 
On the other hand, if you disbelieved it {\it a priori}: $P$(susy) = 0, then
(\ref{twentytwo}) tells you that you still disbelieve it! It seems to me that one
needs some external rationale to determine the measure in MSSM parameter space
and the corresponding 
$P({\rm susy})$. Fine tuning in the sense (\ref{twentyone})  is one proposal, but
there may be better ones.

The new LEP constraints on MSSM particle masses were reviewed
here~\cite{McPherson}. Among the most important ones is
\beq
m_{\chi^\pm} > 95~{\rm to}~ 90 ~{\rm GeV}
\label{twentythree}
\eeq
except in the deep Higgsino region. Combining with other searches, including
associated neutralino production $e^+e^-\rightarrow \chi\chi^\prime$, one
now finds
\beq
m_\chi > 32 ~{\rm GeV}
\label{twentyfour}
\eeq
for all $m_0$ and $\tan\beta$. If $\tilde t \rightarrow  c\chi$ decays dominate
and $m_{\tilde t} -m_\chi \leq$ 10 GeV, one has
\beq
m_{\tilde t} > 87 ~{\rm GeV}
\label{twentyfive}
\eeq
In addition
\beq
m_{\tilde \ell} > 88 ~{\rm GeV}
\label{twentysix}
\eeq
for $m_{\tilde\ell} - m_\chi \gappeq$ 5 GeV, with somewhat weaker limits for
$m_{\tilde\mu}$ and $m_{\tilde\tau}$. In addition to sparticle limits, very
important constraints on the MSSM parameter space come from the Higgs
limit~\cite{Felcini}: the
best limit from an individual LEP experiment is
\beq
m_H > 95.2 ~{\rm GeV}
\label{twentyseven}
\eeq
and the joint sensitivity is to $m_H \lappeq$ 98 GeV. A few candidate events have
been reported~\cite{Felcini}, but there is no significant signal yet.

One may analyze the constraints these limits impose jointly in the space of MSSM
parameters. I should like to underline the importance of including radiative
corrections to the relations between these parameters and physical sparticle
masses~\cite{EFGOS}. These are included routinely for Higgs masses, but
not for chargino and neutralino masses, whereas these
are also significant. In particular, the change in
sensitivity in the $(\mu , M_2)$ plane when these radiative corrections are
included is comparable to the change in physics reach between one year's run of
LEP and the next, as the beam in energy is increased, so these should not 
be neglected in a complete analysis of LEP data~\cite{EFGOS}.

Fig.~4 shows the impacts of various LEP  constraints in the $(m_{1/2}, m_0)$
plane~\cite{EFOSi}, including also the requirement that the electroweak
vacuum be stable
against transitions to vacua that violate charge and/or colour
conservation (CCB)~\cite{AF}.
These are weakest for $A \simeq -m_{1/2}$ as plotted in Fig.~4, but even so
they indicate a preference for
\beq
110 ~{\rm GeV} \lappeq m_{1/2} \lappeq 400 ~{\rm GeV},~~~
80 ~{\rm GeV} \lappeq m_0 \lappeq 170 ~{\rm GeV}
\label{twentyeight}
\eeq
when combined with the cosmological relic density limit. However, it is not clear
whether vacuum stability is an absolute necessity. It might be sufficient if the
early Universe evolved into our vacuum, and if this is metastable with a lifetime
$\gappeq 10^{10}$ years.

\begin{figure}
\hglue1.5cm
\epsfig{figure=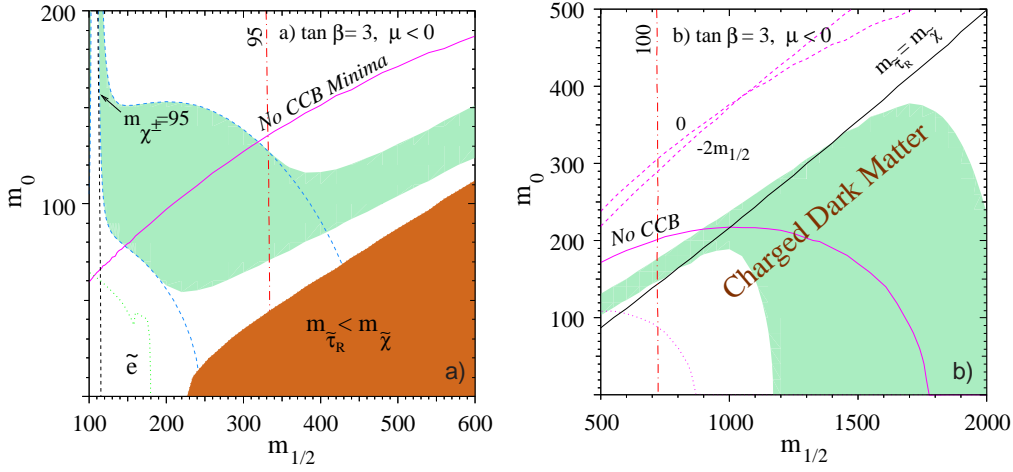,width=13.5cm}
\label{FIG4}
\caption[]{\it Experimental, theoretical and cosmological constraints in
the
$(m_{1/2}, m_0)$ plane, for tan$\beta = 3, \mu < 0$, 
indicating the physics
reach of LEP for $\tilde e, \chi^\pm$ and Higgs searches~\cite{EFOSi}. The light shaded
region is that where $0.1 < \Omega_\chi h^2 < 0.3$ after including
coannihilation effects~\cite{EFO}, and the region plagued
by charge and colour-breaking (CCB) minima is also delineated. Panel (a)
is extended in panel (b) to larger values of $m_{1/2}$, showing how the
upper limit $m_\chi < 600$~GeV may be reached.}
\end{figure}

The cosmological domains in Fig.~4 include the effects of coannihilation 
between the neutralino LSP and the next-to-lightest supersymmetric particle
(NLSP), which can be important if their mass difference $\Delta m \lappeq
{\cal O}(m_\chi / 10)$~\cite{EFO}. In the region of Fig.~4, the NLSP is
the
$\tilde\tau_R$, and
coannihilations with the $\tilde\mu_R$ and $\tilde e_R$ are also significant,
particularly because the $\chi\chi$ annihilations are suppressed in the
non-relativistic limit. The coannihilations suppress the relic density for large
$m_{1/2}$, so that the LSP may be substantially heavier than was previously
thought. Combining all the new LEP constraints~\cite{EFOSi}, we may
estimate that
\beq
m_\chi \gappeq  50 ~{\rm GeV},~~~
\tan\beta\gappeq 1.8
\label{twentynine}
\eeq
and
\beq
m_\chi \lappeq 600 ~{\rm GeV}
\label{thirty}
\eeq
when coannihilations are taken into account.

The lower limit (\ref{twentynine})
is tantalizingly close to the report from the DAMA
collaboration~\cite{DAMA}, in which they
fail to exclude the presence of an annual modulation of a recoil signal, which
could correspond to
$m_\chi
\sim$ 60 GeV with large errors. Not only is the indicated mass compatible with
the experimental limit (\ref{twentynine}), but also the corresponding
cross section is compatible with some theoretical models, particularly if the
MSSM Higgs is light~\cite{Bottino}. However, care should be exercised in
interpreting the
possible annual modulation: we shall need to see data over a complete annual
cycle and be reassured that any modulation could not be due to some
non-fundamental seasonal effect.

There was considerable discussion here of possible phenomenological signatures of
large extra space dimensions~\cite{LykkenD}. In ``classical" perturbative
string theory, the
gauge and gravitational interactions are unified at an energy $E = m_s$ where
they become equal:
\beq
{e^2\over r} = {E^2\over m^2_P}~~ {1\over r}
\label{thirtyone}
\eeq
corresponding to a string scale $m_S = {\cal O}(\sqrt{\alpha}) m_P \simeq
10^{18}$ GeV,
at which six extra dimensions appear. This is modified in
``post-classical" $M$
theory, where one can enforce $m_s \simeq m_{GUT}\simeq 10^{16}$ GeV as indicated
by LEP, by postulating an extra dimension with characteristic size $R \gg
m^{-1}_{GUT}$~\cite{HW}. Above this scale, the Newton potential is
modified:
\beq
{E^2\over m^2_P}~~ {1\over r} \rightarrow  {E^3\over \tilde m^3_P}~~ {1\over r}
\label{thirtytwo}
\eeq
whereas the gauge interactions do not feel the extra dimension. The resulting
picture of space time resembles a pair of capacitor plates.
There are two infinitely large four-dimensional plates separated by a distance $R
\gg m^{-1}_{GUT}$, and each ``point" in this picture can in fact be resolved into
six compactified dimensions at energies $E \gappeq m_{GUT}$. 

In the new ``deconstructionist" approach~\cite{Dvali,Donini}, one asks the
natural and important
question: how large could the extra dimension(s) be? In particular, could the
scale of gravity be as low as 1 TeV, in which case the hierarchy problem is
entirely reformulated. If $n$ extra dimensions appear at a scale $R$ so that the
Newton potential $\propto ~~ 1/r^{1+n}$ at shorter distances, one has
\beq
m^2_P = m_s^{n+2}~R^n
\label{thirtythree}
\eeq
and if $m_s \simeq$ 1 TeV by construction, $R \simeq 10^{13}$ m (1 mm) (10 fm)
for $n$ = 1(2)(6). The Newton $1/r$ law has been checked down to $r \gappeq$ 1
cm, so one needs $n \gappeq 2$ in (\ref{thirtythree}): near-future experiments
may be able to probe down to $r \simeq 10 \mu m$. Some of the most
important constraints on such theories come from astrophysics and
cosmology~\cite{disastro}, including graviton
emission from supernovae, the cosmic microwave background, Big Bang
nucleosynthesis and inflation, which
appear to indicate that $m_s$ may need to
be considerably larger than 1 TeV.

There has been some discussion whether gauge interactions might also
``feel"
(some of) the bulk dimensions appearing at distances $\lappeq
R$~\cite{DDG}. This is not the
case in ``canonical" string or $M$ theory, and is actually impossible in the
standard formulation of five-dimensional supergravity~\cite{FiveD}. It may
be possible to
mimic the very successful GUT prediction for $\sin^2\theta_W$ in 
such a scenario~\cite{DDG},
but I need reassurance how naturally and precisely this may work.
Among the signatures for such higher-dimensional theories discussed
here~\cite{Dvali} are
missing-energy events: $e^+e^- \rightarrow \gamma$ + (gravitons), $gg \rightarrow
g$ + (gravitons), and angular distortions in $e^+e^-\rightarrow\mu^+\mu^-$ due to
higher-dimensional graviton exchange. The latter has been used to show that $m_s
\gappeq$ 0.6 TeV for $n = 2$ on the basis of LEP data~\cite{Holt}. Among
the many issues being studied in such models
are flavour-changing processes, baryon stability and neutrino masses. 

An interesting theoretical question for such models is whether supersymmetry is
still necessary. Not for solving the hierarchy problem in the traditional way,
but string theory still (apparently) requires supersymmetry for its consistency,
so it should still appear around $m_s (\sim$ 1 TeV?). The traditional hierarchy
problem is now repackaged as the question why $R$ is so large. Since the fixing
of the compactification radius and related moduli are not understood in
conventional string theory, though, do we have any good reason to think that $R$
could not be large?

\section{Future Projects}

There were many discussions here of the future prospects in electroweak projects,
of which I give here a few highlights. Over the next two years, 1999 and 2000,
LEP should finally reach its design energy of 200 GeV in the centre of mass,
perhaps even a bit more. Clearly this provides further opportunities to search
for supersymmetric particles, etc., but the most exciting prospects may be those
for the Higgs boson search. We see in Fig.~1
that the highest probability
density for its mass is the range just at the present direct search limit,
and within the reach $m_H \lappeq$ 110 GeV of LEP 200+$\epsilon$. This is,
moreover, the most likely range in the MSSM~\cite{MSSMH}.

If LEP does not find the Higgs, it may be found at the Fermilab Tevatron collider
during one of its future runs, starting in 2000. The
Tevatron collider may also be able to find, or at least exclude, MSSM Higgs
bosons over a considerable range of parameter space~\cite{Carena}. The
reach in the space of MSSM soft supersymmetry breaking parameters will also be
extended at the Tevatron collider~\cite{BK}. Overall, there is
significant chance that the Tevatron collider will be able to steal some of the
LHC's supersymmetric thunder.

Turning now to neutrino physics, there is an extensive programme of work for
long-baseline experiments~\cite{Campanelli}. Accelerators produce intense
beams with controlled
energy spectra, fluxes and flavours of neutrinos, that can be monitored and
adjusted. These may be needed to convince sceptics of the reality of neutrino
oscillations, and in order to make precision measurements. Among the key
measurements to make are those of $\nu_\mu$ disappearance and the neutral- to
charged-current ratio, involving the comparison of rates in near and far
detectors, as planned for the K2K~\cite{K2K} and MINOS~\cite{MINOS} 
experiments. Another key measurement
is that of $\nu_e$ appearance, which is not expected to be the dominant
mode of oscillation, but might be present below the 10 \% level.

In my view, the key experiment to pin down the interpretation of the atmospheric
neutrino data will be the search for $\tau$ appearance
subsequent to $\nu_\mu \rightarrow \nu_\tau$ oscillations. This is the
favoured interpretation of the current data,
but on the basis of indirect arguments. I believe there is no substitute for
direct observation of $\tau$ production in a long-baseline experiment. The
interpretation would be unambiguous, since the accelerator and other backgrounds
are negligible. Only in this way could alternative interpretations such as
$\nu_\mu$ oscillations into sterile neutrinos be definitively rejected. Remember
Jimmy Hoffa, and the basic legal principle ``if you have not seen the body, you
have not proven there was a murder". Remember also the gluon: there were many
indirect and theoretical arguments that it should exist, and that it 
should have
unit spin, but everybody remembers the direct observation of gluon jets in the
reaction $e^+e^-\rightarrow \bar qqg$~\cite{EGR} as the ``discovery of the
gluon"~\cite{TASSO}.

Three long-baseline neutrino oscillation projects have been approved so
far~\cite{Campanelli}. Two
are in Japan: KamLAND, which sets out to detect $\nu_e$ from power reactors and
should be able to confirm or refute the LMA solution to the solar neutrino
problem, and K2K, which is a $\nu_\mu$ disappearance experiment using a KEK beam
that can probe aboout a half of the region in $(\sin^2 2\theta , \Delta m^2)$
favoured by Super-Kamiokande. KamLAND should start taking data in 2001, and K2K
has already started. The only approved project with a beam energy high enough to
detect $\nu_\tau$ via $\tau$ production is NuMI. The MINOS detector (to
start data-taking in 2002, with the detector to be
completed in 2003) should be able to cover all the Super-Kamiokande region with a
$\nu_\mu$ disappearance measurement, but has a limited ability to detect $\tau$
leptons. A proposal has now been made to add to MINOS a $\tau$ detection module
based on emulsion tracking, which could be completed in
2004.

In Europe, CERN and the Gran Sasso laboratory are proposing a slightly higher
energy beam optimized for $\tau$ production, that could be ready in
2005~\cite{NGS}. It
would make optimized use of the CERN infrastructure, including one of the
lines
transferring protons to the LHC, and benefit from the fact that the Gran
Sasso
experimental halls were, from the beginning, oriented so as to profit maximally
from a CERN neutrino beam. Experimental concepts now being
developed~\cite{tauapp} should be
able comfortably to detect $\tau$ leptons over all the parameter range allowed
by Super-Kamiokande, as seen in Fig.~3.

The LHC physics programme was not much discussed at this meeting. Paramount is
the search for the ``holy Higgs". This has been well studied in the Standard
Model, and extensive studies have been made in the MSSM~\cite{Was}, but
the latter has not
been completely explored. LEP studies have revealed the possibility of a
reduction in the $h\rightarrow \bar bb$ decay mode that usually dominates. The
possibilities of  interference effects in $\sigma (gg\rightarrow h)$ and $\Gamma
( h\rightarrow
\gamma\gamma)$ merit further work, and this is one area where more dialogue
between theorists and experimentalists preparing LHC detectors would be useful.
Moriond could play a useful role in this regard, particularly in encouraging and
motivating the young people who will do most of the work. The success of the LHC
will require a big, coherent effort from all sectors of the community.

Also important at the LHC will be the search for supersymmetric particles. Here
the production cross sections are well understood, and potentially complicated
decay chains such as $\tilde g \rightarrow \tilde b \bar b$, $\tilde b \rightarrow
\chi_2 b,
\chi_2 \rightarrow \chi \ell^+\ell^-$ have been
explored. The LHC will be able
to explore $m_{\tilde g,\tilde q} \lappeq$ 2 TeV and $m_\ell \lappeq$ 400 GeV,
and could make some detailed studies of sparticle
spectroscopy~\cite{Hinchliffe}. It used to be
thought that the entire region of MSSM parameter space favoured by cosmology
could be covered comfortably by the LHC, several times
over~\cite{Abdullin}. However, the
coannihilation effects~\cite{EFO,EFOSi} mentioned earlier stretch the
cosmologically favoured
region up to larger $m_{1/2}$ as seen in Fig.~4, close to 
the LHC reach, so this issue needs to be reviewed.

Beyond the LHC, the physics programme for a linear $e^+e^-$ collider is currently
being extensively studied, particularly the relative advantage of high
luminosities, e.g., with TESLA~\cite{Schulte}. One example is the study of
Standard Model Higgs
decays: if the Higgs boson has a mass in the range favoured by precision
electroweak measurements, its $\bar bb, \bar cc, gg$ and $\tau^+\tau^-$ branching
ratios could be measured accurately. However, the total Higgs decay width could
not be measured without implementing the $\gamma\gamma$ collider option, which is
not currently part of the TESLA baseline.

I close by mentioning the most futuristic project idea discussed here: muon
storage rings. These could be developed in a three-step
scenario~\cite{Yellow}, starting with a
neutrino factory~\cite{Dydak} using the known fluxes, flavours,
charges and spectra from muon
decay - the ``ultimate weapon" for $\nu$ oscillation
studies~\cite{Gavela}, continuing with one
or more $\mu^+\mu^-$ colllider Higgs factories suitable for measuring the total
decay width(s), reducing the allowed phase space of the MSSM and perhaps
providing a new window on CP violation -- the ``ultimate weapon" for Higgs
studies~\cite{Janot}, and a high-energy $\mu^+\mu^-$ collider, whose
maximum energy is
currently limited by the potential neutrino-induced radiation
hazard~\cite{Yellow}. 

Very many basic technical issues need to be resolved before the feasibility of
such storage rings and colliders can be judged, but their physics is very
enticing. In the presence of three different masses, neutrinos would have three
real mixing angles and three phases (though two of the latter cannot be measured
at energies $E \gg m_\nu$). Thus there is a programme of work as rich as that for
the B factories now coming on line~\cite{Gavela}. As discussed
here~\cite{Gavela} and reflected in Fig.~5, the following are the
sensitivities of appearance and disappearance experiments as functions of the
beam length and energy:
\bea
\Delta m^2_{ij} &:& E_\mu^{-1/2}~,~~L^{-1/2}~E_\mu^{1/4} \nonumber \\
&&\nonumber \\
\sin^2\theta_{ij} &:& LE_\mu^{-3/2}~,~~L^{1/2}~E_\mu^{-3/4}
\label{thirtyfour}
\eea
These dependences imply that very-long-baseline experiments $(L \gappeq$ 3000 km)
are not necessarily the best. As shown here, long-baseline experiments have very
interesting physics reaches for oscillation parameters in a three-generation
analysis. In addition, there is a chance of observing CP-violating effects,
particularly with a very long baseline~\cite{Gavela}.

\begin{figure}
\hglue2.5cm
\epsfig{figure=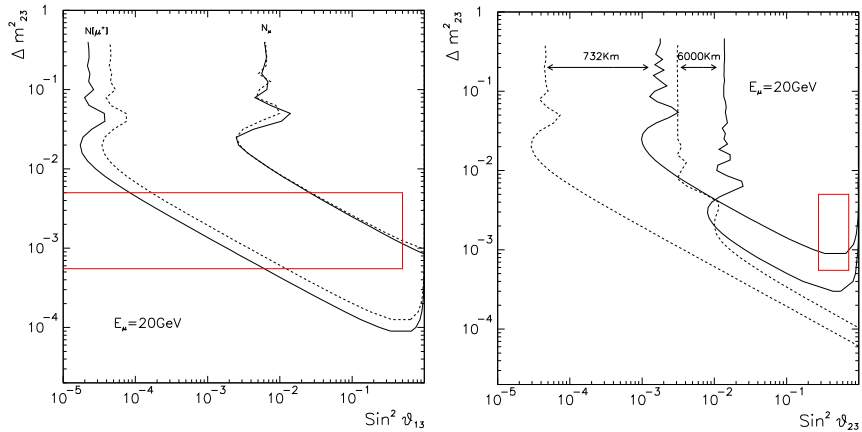,width=12cm}
\label{FIG5}
\caption[]{\it The sensitivities of long-baseline neutrino experiments
using beams from a muon storage ring used as a neutrino
factory~\cite{Gavela}: (a)
to search for mixing between the first- and third-generation
neutrinos via appearance (left lines) and disappearance (right lines) for
$\theta_{23} = 45^o$ (solid lines) and $30^o$ (dashed lines), 
assuming a baseline of 732~km, and (b)
to search for mixing between the second- and third-generation neutrinos
via appearance (dashed lines) and disappearance (solid lines), assuming 
the indicated beam lengths. The boxes represent current indications and
limits.}
\end{figure}

A first Higgs factory with $E_{cm} \sim$ 110 GeV  could measure the mass of the
Higgs with a precision of 0.1 MeV~\cite{Yellow,Janot}. In the MSSM, a
second Higgs factory with
$E_{cm} \sim m_{H,A}$ could measure their masses with a precision of 10 MeV and
their (relatively large) widths with a precision of 50 MeV. The twin peaks may
also offer prospects for observing CP-violating observables. Such $\mu^+\mu^-$
colliders offer precision tests of the MSSM~\cite{Yellow,Janot}.

Finally, the Table summarizes some of the comparative physics strengths of the
LHC and high-energy $e^+e^-$ and $\mu^+\mu^-$ colliders~\cite{Yellow}.
Each has unique
advantages (e.g., flavour non-universality and energy resolution in the case of a 
$\mu^+\mu^-$ collider), and a key role to play in the future elucidation of
electroweak physics. The LHC is already on the way, and we hope that a
first-generation $e^+e^-$ linear collider can follow it. In the mean time, let us
work to make $\mu^+\mu^-$ colliders realistic options for the more distant future.

\begin{table}
\caption{{\it Accessibilities of various possible new physics phenomena
with 
the LHC, a 4-TeV $e^+e^-$ collider and a 4-TeV $\mu^+\mu^-$ collider. The
crosses X denote inaccessible features of models, the symbols Y denote
accessible features. We indicate by F (E) the topics where flavour
non-universality (energy resolution) is a particular advantage for a 
$\mu^+\mu^-$ collider, and by $\gamma$ (P) topics where $\gamma\gamma$
collisions (polarization) confer advantage on an $e^+e^-$ collider.}}
\vspace{0.8cm}
\label{Table1}
\begin{center}
\begin{tabular}{|l|l|l|l|}\hline
Physics topic & LHC & $e^+e^-$ & $\mu^+\mu^-$ \\ \hline
Supersymmetry &&& \\ 
~~~Heavy Higgses H, A & X ? & ?: $\gamma$ & Y: F,E \\
~~~Sfermions & $\tilde q$ & $\tilde\ell$ & $\tilde \ell$ : F \\
~~~Charginos & X ? & Y: P & Y: F,E \\
~~~$R$ violation & $\tilde q$ decays & $\lambda_{1ij}$ & $\lambda_{2ij}$:F,E \\
~~~SUSY breaking & some & more & detail: F E \\ \hline
Strong Higgs sector &&& \\
~~~Continuum  &$\leq$ 1.5 TeV & $\leq$ 2 TeV & $\leq$ 2 TeV \\
~~~Resonances & scalar, vector& vector, scalar & vector (E), scalar (F) \\
\hline
Extra dimensions &&& \\
~~~Missing energy & large $E_T$ & Y & Y: E? \\
~~~Resonances & $q^*, g^*$ & $\gamma^*, Z^*, e^*$ & $\gamma^*, Z^*, \mu^*$: E
\\ \hline
\end{tabular}
\end{center}
\end{table}


\vspace{3cm}



















\begin{thebibliography}{99}
\bibitem{LEPEWWG} LEP Electroweak Working Group, CERN preprint EP/99-15;
updates may be found at \\
{\tt http://www.cern.ch/LEPEWWG/Welcome.html}.

\bibitem{SLD} N. de Groot, SLD collaboration, talk at
this meeting.

\bibitem{BW} S.C. Bennett and C.E. Wieman,
Phys. Rev. Lett. {\bf 82}, 2484 (1999).

\bibitem{Casal}
See also: R. Casalbuoni, S. De Curtis, D. Dominici and R. Gatto,
hep-ph/9905568.

\bibitem{Lanc} M. Lancaster, talk at this meeting; for other Tevatron
measurements, see N. Graf, talk at this meeting.

\bibitem{Riu} I. Riu, talk at this meeting.

\bibitem{Hapke} M. Hapke, talk at this meeting.

\bibitem{EG} J. Ellis and K. Geiger,
Phys. Lett. {\bf B404}, 230 (1997).

\bibitem{GEHW} K. Geiger, J. Ellis, U. Heinz and U. Wiedemann,
hep-ph/9811270.

\bibitem{Felcini} M. Felcini, talk at this meeting.

\bibitem{Davier} M. Davier and A. Hoecker,
Phys. Lett. {\bf B419}, 419 (1998).

\bibitem{LykkenTeV} J. Lykken, opening talk at this meeting, see also: \\
{\tt http://www.fnal.gov/pub/hep{\_}descript.html}.

\bibitem{Qi} X.-R. Qi, talk at this meeting.

\bibitem{g2} E. Perazzi, talk at this meeting;
D. Urner, talk at this meeting.

\bibitem{invalid} T. Hambye and K. Riesselmann,
hep-ph/9708416.

\bibitem{superweak} L. Wolfenstein
Phys. Rev. Lett. {\bf 13}, 562 (1964).

\bibitem{KTeV} E. Blucher, KTeV Collaboration, talk at this meeting; 
see also: A. Alavi-Harati et al., KTeV Collaboration, hep-ex/9905060.

\bibitem{KM} M. Kobayashi and K. Maskawa,
Prog. Theor. Phys. {\bf 49}, 652 (1973).

\bibitem{EGN} J. Ellis, M.K. Gaillard and D.V. Nanopoulos,
Nucl. Phys. {\bf B109}, 213 (1976).

\bibitem{Wagner} C.E.M. Wagner, talk at this meeting;
see also: A. Pilaftsis and C.E.M. Wagner, hep-ph/9902371.



\bibitem{Kalinowski} J. Kalinowski, talk at this meeting; see also:
B.~Grzadkowski, J.F.~Gunion and J.~Kalinowski,
hep-ph/9902308.



\bibitem{EGNR} J. Ellis, M.K. Gaillard, D.V. Nanopoulos and S. Rudaz,
Phys. Lett. {\bf B131}, 285 (1977).

\bibitem{NA31} G.D. Barr et al., NA31 Collaboration,
Phys. Lett. {\bf B317}, 233 (1993).

\bibitem{GW} F. Gilman and M. Wise,
Phys. Lett. {\bf B83}, 83 (1979).

\bibitem{FR} J.M. Flynn and L. Randall,
Phys. Lett. {\bf B224}, 221 (1989).

\bibitem{latest} 
Y.~Keum, U.~Nierste and A.I.~Sanda,
hep-ph/9903230;
S.~Bosch, A.J.~Buras, M.~Gorbahn, S.~Jager, M.~Jamin, M.E.~Lautenbacher
and L.~Silvestrini,
hep-ph/9904408.

\bibitem{Lellouch} L. Lellouch, talk at this meeting.

\bibitem{Videau} H. Videau, talk at this meeting.

\bibitem{sepsilon} A. Masiero and H. Murayama, hep-ph/9903363;
K.S. Babu, B. Dutta and R.N. Mohapatra, hep-ph/9905464.

\bibitem{Silvestrini} L. Silvestrini, talk at this meeting,
hep-ph/9906202.

\bibitem{Mikulec} I. Mikulec, NA48 Collaboration, talk at this meeting.

\bibitem{CPTQG} J.~Ellis, J.S.~Hagelin, D.V.~Nanopoulos and M.~Srednicki,
Nucl. Phys. {\bf B241}, 381 (1984);
J.~Ellis, J.L.~Lopez, N.E.~Mavromatos and D.V.~Nanopoulos,
Phys. Rev. {\bf D53}, 3846 (1996);
R.~Adler {\it et al.},
CPLEAR Collaboration,
Phys. Lett. {\bf B364}, 239 (1995).

\bibitem{CPLEAR} 
A.~Angelopoulos {\it et al.},
CPLEAR Collaboration,
Phys. Lett. {\bf B444}, 43 (1998).

\bibitem{Filipcic} A. Filipcic, CPLEAR Collaboration, talk at this
meeting;

\bibitem{EM} J. Ellis and N.E. Mavromatos,
hep-ph/9903386.

\bibitem{Lola} S. Lola, talk at this meeting; see also:
L.~Alvarez-Gaum\'e, C.~Kounnas, S.~Lola and P.~Pavlopoulos,
hep-ph/9812326
and hep-ph/9903458.

\bibitem{Yamanaki} T. Yamanaki, KTeV Collaboration, talk at this meeting;
see also: \\
{\tt http://www.fnal.gov/pub/hep{\_}descript.html}.

\bibitem{Schmidt} M. Schmidt, CDF Collaboration, talk at this meeting;
see also: CDF Collaboration, CDF/PUB/BOTTOM/CDF/4855 (1999).

\bibitem{Romanino} A. Romanino, talk at this meeting; see also:
R. Barbieri, L.J. Hall and A. Romanino, hep-ph/9812384.

\bibitem{Ali} 
A.~Ali and D.~London,
hep-ph/9903535.

\bibitem{Moore} T. Moore, SLD Collaboration, talk at this meeting.

\bibitem{Bloch} D. Bloch, talk at this meeting.

\bibitem{BEG} R. Barbieri, J. Ellis and M.K. Gaillard,
Phys. Lett. {\bf 90B}, 249 (1980).


\bibitem{seesaw} 
M. Gell-Mann, P. Ramond and R. Slansky, Proceedings of the
Stony Brook Supergravity Workshop, New York, 1979, eds. P. Van
Nieuwenhuizen and D. Freedman (North-Holland, Amsterdam);
T. Yanagida, Proceedings of
the  Workshop  on Unified  Theories  and  Baryon  Number in the
Universe,  Tsukuba,  Japan 1979, eds. A.  Sawada and A.
Sugamoto, KEK Report No.  79-18.

\bibitem{Allanach} B. Allanach, talk at this meeting; see also:
B. Allanach, Phys. Lett. {\bf B450}, 182 (1999).

\bibitem{ELLN} For one particular take on this, see:
J. Ellis, G. Leontaris, S. Lola and D.V. Nanopoulos,
hep-ph/9808251; for a recent review, see:
G.~Altarelli and F.~Feruglio,
hep-ph/9905536.


\bibitem{Haxton} W. Haxton,
Prog. Part. Nucl. Phys. {\bf 40}, 101 (1998).

\bibitem{nofit}
G.L.~Fogli, E.~Lisi, A.~Marrone and G.~Scioscia,
Phys. Rev. {\bf D59}, 117303 (1999)
and
hep-ph/9904248;
see also: S. Pakvasa, hep-ph/9905426.

\bibitem{Mills} G. Mills, LSND Collaboration, talk at this meeting.

\bibitem{Jannakos} T. Jannakos, KARMEN Collaboration, talk at this
meeting.

\bibitem{Casper} D. Casper, Super-Kamiokande Collaboration, talk at this
meeting; see also: K. Scholberg, hep-ex/9905016.

\bibitem{SuperK}
Y. Fukuda et al., Super-Kamiokande collaboration,
Phys. Rev. Lett. {\bf 81}, 1562 (1998). 

\bibitem{EW}
T.~Futagami et al.,
Super-Kamiokande Collaboration,
astro-ph/9901139.

\bibitem{MACRO} D. Michael, MACRO Collaboration, talk at this meeting; see
also:
R. Ronga, MACRO Collaboration, hep-ex/9905025;
A. Surdo, MACRO Collaboration, hep-ex/9905028.

\bibitem{Soudan}
W.W.M. Allison et al., Soudan II Collaboration, Phys. Lett. {\bf B449},
137 (1999).

\bibitem{Lipari} P. Lipari, hep-ph/9905506.

\bibitem{balloon}
R. Bellotti et al., hep-ex/9905012.

\bibitem{Marteau} J. Marteau, talk at this meeting.

\bibitem{MSW} L. Wolfenstein, Phys. Rev. {\bf D17}, 2369 (1978);
S.P. Mikheev and A.Y. Smirnov, 
Sov. J. Nucl. Phys. {\bf B42}, 913 (1985) and Nuov. Cim. {\bf 9C}, 17
(1986).

\bibitem{Smirnov} A.Y. Smirnov, talk at this meeting.

\bibitem{Nicolo} D. Nicolo, Chooz Collaboration, talk at this meeting.

\bibitem{Wang} Y. Wang, Palo Verde Collaboration, talk at this meeting.

\bibitem{bighep} J. N. Bahcall and P.I. Krastev, Phys. Lett. {\bf B436},
243 (1998).

\bibitem{nadir} J.N. Bahcall, P.I. Krastev and A.Y. Smirnov,
hep-ph/9905220 and
references therein.

\bibitem{cosmo} R.A.C. Croft, W. Hu and R. Dav\'e, astro-ph/9903335
and references therein.

\bibitem{HET} W. Hu, D.J. Eisenstein and M. Tegmark,
Phys. Rev. Lett. {\bf 80}, 5255 (1998).

\bibitem{Messina} M. Messina, CHORUS Collaboration, talk at this meeting.

\bibitem{Salvatore} P. Salvatore, NOMAD Collaboration, talk at this
meeting.

\bibitem{Klap} L. Baudis et al., hep-ex/9902014.

\bibitem{EL}
J.~Ellis and S.~Lola,
hep-ph/9904279.

\bibitem{others} For another viewpoint, see:
J.A. Casas, J.R. Espinosa, A. Ibarra and I. Navarro,
hep-ph/9904395 and hep-ph/9905381.

\bibitem{Matias} J. Matias, talk at this meeting, hep-ph/9905380; see
also:
A. Abada, J. Matias and R. Pittau, Phys. Rev. {\bf D59}, 013008 (1999),
Nucl. Phys. {\bf B543}, 255 (1999) and hep-ph/9809418.


\bibitem{Tytgat} M. Tytgat, talk at this meeting; see also:
K. Kiers and M. Tytgat, hep-ph/9905532.

\bibitem{Jack} I. Jack and D.R.T. Jones, hep-ph/9903365.

\bibitem{Arnoud} Y. Arnoud, talk at this meeting.

\bibitem{EHNOS} 
J.~Ellis, J.S.~Hagelin, D.V.~Nanopoulos, K.~Olive and M.~Srednicki,
Nucl. Phys. {\bf B238}, 453 (1984).

\bibitem{GiuRat}
G.F. Giudice and R. Rattazzi, hep-ph/9801271.

\bibitem{CDF}
F. Abe et al., CDF Collaboration, Phys. Rev. Lett. {\bf 81} (1998) 1791
and
Phys. Rev. {\bf D59}, 092002 (1999).



\bibitem{Antonelli} M. Antonelli, talk at this meeting;


\bibitem{Numerotski} A. Numerotski, talk at this meeting;
other Tevatron limits were reported here by D. Claes, talk at this
meeting.

\bibitem{EFOS} 
J. Ellis, T. Falk, K.A. Olive and M. Schmitt,
Phys. Lett. {\bf B388}, 97 (1996) and {\bf B413}, 355 (1997).

\bibitem{price} J. Ellis, K. Enqvist, D.V. Nanopoulos and F. Zwirner,
Mod. Phys. Lett. {\bf A1}, 57 (1986);
R. Barbieri and G.F. Giudice, Nucl.
Phys. {\bf B306}, 63 (1988).

\bibitem{CEOP} 
P.H. Chankowski, J. Ellis and S. Pokorski,
Phys. Lett. {\bf B423}, 327 (1998);
R. Barbieri and A. Strumia, Phys. Lett. {\bf B433}, 63 (1998);
P.H. Chankowski, J. Ellis, M. Olechowski and S. Pokorski,
Nucl. Phys. {\bf B544}, 39 (1999).

\bibitem{Strumia} A. Strumia, talk at this meeting, hep-ph/9904247.

\bibitem{McPherson} R. McPherson, talk at this meeting.

\bibitem{EFGOS} J. Ellis, T. Falk, G. Ganis, K.A. Olive and M. Schmitt,
Phys. Rev. {\bf D58}, 095002 (1998).

\bibitem{EFOSi} J. Ellis, T. Falk, K.A. Olive and M. Srednicki,
hep-ph/9905481.

\bibitem{AF} S. Abel and T. Falk, Phys. Lett. {\bf B444}, 427 (1998).

\bibitem{EFO} J. Ellis, T. Falk and K.A. Olive,
Phys. Lett. {\bf B444}, 367 (1998).

\bibitem{DAMA} R. Bernabei et al., DAMA Collaboration,
Phys. Lett. {\bf B450}, 448 (1999).

\bibitem{Bottino} A. Bottino, F. Donato, N. Fornengo and S. Scopel,
Phys. rev. {\bf D59}, 095003, 095004 (1999).

\bibitem{LykkenD} J. Lykken, talk at this meeting.


\bibitem{HW} 
P. Horava and E. Witten, Nucl. Phys. {\bf B460}, 506 (1996);
E. Witten, Nucl. Phys. {\bf B471}, 135 (1996);
P. Horava and E. Witten, Nucl. Phys. {\bf B475}, 94 (1996).


\bibitem{Dvali} G. Dvali, talk at this meeting.

\bibitem{Donini} A. Donini, talk at this meeting.

\bibitem{disastro} N. Arkani-Hamed, S. Dimopoulos and G. Dvali,
Phys. Rev. {\bf D59}, 086004 (1999);
S. Cullen and M. Perelstein, hep-ph/9903422;
L.J. Hall and D. Smith, hep-ph/9904267;
V. Barger, T. Han, C. Kao and R.-J. Zhang,
hep-ph/9905474.

\bibitem{DDG} K. Dienes, E. Dudas and T. Gherghetta,
Phys. Lett. {\bf B436}, 55 (1998) and Nucl. Phys. {\bf B537}, 47 (1999).

\bibitem{FiveD} G. Sierra, Phys. Lett. {\bf 154B}, 379 (1985);
M. Gunaydin, G. Sierra and P.K. Townsend, Nucl. Phys. {\bf B355}, 573
(1985);
I. Antoniadis, S. Ferrara and T. Taylor,
Nucl. Phys. {\bf B460}, 489 (1996);
A. Lukas, B.A. Ovrut, K.S. Stelle and D. Waldram,
Phys. Rev. {\bf D59}, 086001 (1999) and hep-th/9806051;
J. Ellis, Z. Lalak, S. Pokorski and W. Pokorski, Nucl. Phys. {\bf B540},
149 (1999);
J. Ellis, Z. Lalak and W. Pokorski, hep-th/9811133.


\bibitem{Holt} P.J. Holt, talk at this meeting.

\bibitem{MSSMH} M. Carena, M. Quiros and C.E.M. Wagner,
Nucl. Phys. {\bf B461}, 407 (1996);
H.E. Haber, R. Hempfling and A.H. Hoang, Zeit. f\"ur Phys. {\bf C75}, 539
(1997).

\bibitem{Carena} M. Carena, talk at this meeting; see also:
M. Carena, S. Mrenna and C.E.M. Wagner, hep-ph/9808312.


\bibitem{BK} V. Barger and C. Kao,
hep-ph/9811489.

\bibitem{Campanelli} M. Campanelli, talk at this meeting; see also:
hep-ex/9905035.

\bibitem{K2K}
Y.~Oyama, K2K Collaboration,
hep-ex/9803014.



\bibitem{MINOS}
E.~Ables et al., MINOS Collaboration,
Fermilab proposal P-875.


\bibitem{EGR} J. Ellis, M.K. Gaillard and G.G. Ross, 
Nucl. Phys. {\bf B111}, 253 (1976).

\bibitem{TASSO} 
R.~Brandelik et al.,
TASSO Collaboration,
Phys. Lett. {\bf 86B}, 243 (1979).


\bibitem{NGS}
G.~Acquistapace et al.,
{\it The CERN neutrino beam to Gran Sasso (NGS): Conceptual technical
design}, CERN-98-02 (1998).



\bibitem{tauapp}
A. Rubbia, et al., ICARUS
Collaboration, {\it A search program of explicit neutrino
oscillations with the ICARUS detector at long distances}, 
CERN-SPSLC-96-58 (1996);
K.~Kodama et al.,
{\it The OPERA tau neutrino appearance experiment in the CERN-Gran Sasso
                  neutrino beam},
CERN-SPSC-98-25 (1998);
M. Doucet, J. Panman and P. Zucchelli, hep-ex/9905029.

\bibitem{Was}
E.~Richter-Was, D.~Froidevaux, F.~Gianotti, L.~Poggioli, D.~Cavalli and
S.~Resconi,
Int. J. Mod. Phys. {\bf A13}, 1371 (1998).

\bibitem{Hinchliffe}
I.~Hinchliffe, F.E.~Paige, M.D.~Shapiro, J.~Soderqvist and W.~Yao,
Phys. Rev. {\bf D55}, 5520 (1997).



\bibitem{Abdullin}
S.~Abdullin and F.~Charles,
Nucl. Phys. {\bf B547}, 60 (1999).



\bibitem{Schulte} D. Schulte, talk at this meeting.

\bibitem{Yellow} 
{\it Prospective Study of Muon Storage Rings at CERN}, eds.
B. Autin, A. Blondel and J.~Ellis, CERN-99-02 (1999).

\bibitem{Dydak} F. Dydak, talk at this meeting.

\bibitem{Gavela} B. Gavela, talk at this meeting; see also:
A. De Rujula, B. Gavela and P. Hernandez,
Nucl. Phys. {\bf B547}, 21 (1999);
M.~Campanelli, A.~Bueno and A.~Rubbia,
hep-ph/9905240.


\bibitem{Janot} P. Janot, talk at this meeting.



\end{thebibliography}
\end{document}